# Gold nanocrescents for remotely measuring and controlling local temperature

Xuan Hoa Vu,[†] Michael Levy,[†] Thomas Barroca,[†] Hong Nhung Tran,[‡] and Emmanuel Fort[†,*]

[†]*Institut Langevin, ESPCI ParisTech, CNRS UMR 7587 & INSERM ERL U979, 1 rue Jussieu, 75238 Paris Cedex 05, France,*

[‡]*Center for Quantum Electronics, Institute of Physics, VAST, 10 Dao Tan, Ba Dinh, Hanoi, Vietnam*

* Corresponding author. E-mail:: emmanuel.fort@espci.fr

## Abstract

We present a novel technique to remotely measure and control the local temperature within a medium. This technique is based on the observation of the rotational Brownian motion of gold nanocrescent particles, which possess a strong anisotropic light interaction due to their plasmonic properties. Rotational scattering correlation spectroscopy performed on a single nanoparticle is able to determine the local temperature with high accuracy. These nano-thermometers can simultaneously play the role of nano-heaters when absorbing the light of a focused laser beam.

The ability to control and measure the local temperature of a medium at the nanometer scale is of substantial value in numerous nanotechnology applications, including nanoelectronics,[1] spectroscopy,[2] nanofluidics,[3–6] nanoscale catalysis[7] and photothermal therapeutic medicine.[8–12] Recently, various strategies have been developed to either perform high-resolution thermal mapping (for example, scanning thermal microscopy,[13] fluorescence polarization anisotropy[14] and fluorescent molecular/polymeric thermometers[15–20]) or remotely control the local temperature using plasmonic[21–24] or magnetic nanoparticles.[25–27] However, none of these techniques can achieve both local temperature sensing and heating.

In this paper, we show that nanoparticles consisting of gold semi-coated dielectric nanobeads, currently called nanocrescents, can simultaneously act as thermal nano-sensors and local nano-heaters that can be remotely activated by light focusing. Such dual functionality is made possible by taking advantage of the strong anisotropic light interaction afforded by their plasmonic resonances.

The local temperature is measured by analyzing the rotational Brownian motion of a single nanocrescent using rotational scattering correlation spectroscopy (RSCS). Because of the particle's anisotropic optical signature, rotational diffusion makes it blink erratically, and the autocorrelation of this signal enables determination of the local temperature of the medium surrounding the particle. Measuring rotational diffusion instead of translational diffusion has many advantages. For example, it can be performed in liquid flows or applied in complex media where translational diffusion is impeded.

The plasmonic properties of gold nanoparticles with nanocrescent shapes have been studied in detail.[3,28–31] They possess plasmonic resonances associated with their anisotropic shape that depend on their relative orientation with the incident light. At these resonant wavelengths, they have large scattering and absorption cross-sections. Their extinction spectra possess two characteristic plasmon resonances associated with an axial and a transverse red-shifted mode. Their strong scattering makes it possible to easily track a single nanoparticle and to deduce its orientation.[32] Because of its strong absorption, an individual particle can be used as an efficient nano-heater generating a local hyperthermia when illuminated.

Here, we provide evidence that nanocrescents can be used to remotely control the heating of their surroundings, and that RSCS is an efficient way to measure the temperature increase. After briefly describing the theoretical model used to deduce the temperature from the experimental intensity autocorrelation functions, we present the experimental setup that we used to measure and control the local temperature at the single-nanoparticle level and discuss the results.

Let us consider a particle of hydrodynamical volume $V_h$ undergoing Brownian motion in a medium of viscosity $\eta(T)$ at temperature $T$. The particle scatters light in the direction of observation $z$ defined in the own local frame of reference of the particle by the angles $\Omega = (\theta, \varphi)$ where $\theta$ is the polar angle and $\varphi$ the azimuth angle. Because of rotational diffusion, $\Omega(t)$ fluctuates with time, and we measure the resulting fluctuation intensity $I(\Omega(t)) \equiv I(t)$.

RSCS consists of analyzing the autocorrelation function $G(\tau) = \langle I(t)I(t+\tau) \rangle$ of this intensity $I(t)$. Here, the brackets denote averaging either over time or over a large number of particles. $G(\tau)$ depends on both the rotational diffusion properties and the geometry of the experimental setup. When stationary, the autocorrelation function $G(\tau)$ can be expressed as:

$$G(\tau) = \iint I(\Omega)I(\Omega')p(\Omega,0;\Omega',\tau)d\Omega d\Omega' \qquad (1)$$

where $p(\Omega,t;\Omega',t')$ is the joint probability density for a nanoparticle to have the direction $\Omega$ at time $t$ and the direction $\Omega'$ at time $t'$.

It is convenient to express $p(\Omega,0;\Omega',\tau)$ as the product $p(\Omega)p(\Omega',\tau|\Omega)$, where $p(\Omega) = 1/4\pi$ is the equiprobability density for a nanoparticle to be on the direction $\Omega$ at any time and $p(\Omega',\tau|\Omega)$ is the conditional probability density for the nanoparticle to be in the direction $\Omega'$ at time $\tau$, knowing that the direction was $\Omega$ at the initial time. The latter probability is, by definition, the standard Green function for the rotational diffusion equation.[33,34] Moreover, for an axially symmetric particle $I(\Omega)$ can be expanded in $Y_{l0}$ spherical harmonics:

$$I(\tau) = \sum_{l=0}^{\infty} c_l Y_{l0}(\Omega) \qquad (2)$$

With $c_l$ the expansion coefficients of $I(\Omega)$ in the orthonormal basis of the spherical harmonics. Inserting this expression and the explicit expression for $p(\Omega',\tau|\Omega)$ in eq. (1), we obtain after some calculations using the addition theorem[35] and the orthogonality relations of the spherical harmonics:

$$G(\tau) = \frac{1}{4\pi} \sum_{l=0}^{\infty} c_l^2 \, e^{-l(l+1)\tau/2\tau_r} \qquad (3)$$

where $\tau_r$ is the Brownian relaxation time given by:

$$\tau_r = \frac{3\eta(T)V_h}{k_B T} \qquad (4)$$

where $k_B$ is the Boltzmann constant.

The angular scattering function $I(\Omega)$, which can be determined experimentally, is the signature of the particle shape and determines the autocorrelation function shape. $G(\tau)$ is a multi-exponential function with the relaxation time $\tau_r$ as the only parameter. According to the theoretical expression of $\tau_r$, one can deduce the local temperature by fitting the experimental autocorrelation intensity function with eq. (3) knowing the hydrodynamical volume $V_h$ and the surrounding viscosity $\eta(T)$. From the general expression of eq. (3), two usual approximations can be performed: cosine emission with $I(\Omega) \propto \cos\theta$ (for all $l \neq 1$, $c_l = 0$), which implies that $G(\tau)$ becomes $G_{l=1}(\tau) = e^{-\tau/\tau_r}$, and the dipolar emission with $I(\Omega) \propto \cos^2\theta$ (for all $l \neq 2$, $c_l = 0$), which implies that $G(\tau)$ becomes $G_{l=2}(\tau) = e^{-3\tau/\tau_r}$.[30]

Nanocrescents are synthetized by a nanosphere lithography technique. A thin layer of metal is deposited on dielectric nanoparticles spin-coated on a glass substrate.[3,32,36] Figure 1a shows a schematic of the principle. For magnetic purification, we use magnetic nanospheres (Estapor, Merck Chimie SAS) consisting of a polystyrene matrix of 140±20 nm that encapsulates 8 nm superparamagnetic iron oxide nanocrystals. A 30-nm-thick gold layer is deposited on a 2-nm-thick adhesion layer of chromium by electron beam evaporation. The nanocrescents are then removed from the glass slide via gentle brushing and sonication in an aqueous solution.[30,31] The obtained dilute nanocrescent colloid is washed by magnetic purification and replacing the supernatant with clean distilled water. Finally, the colloidal solution is concentrated into a small volume of 100% glycerol. We chose glycerol for its high viscosity and strong temperature dependency. Figure 1b shows a transmission electron microscope (TEM) image of an individual nanocrescent and its schematic representation. We can distinguish the gold semi-shell, the polymeric nanosphere and the magnetic nanocrystals embedded in the polymer. TEM images are used to confirm the size distribution of the nanocrescents given in the product specifications.

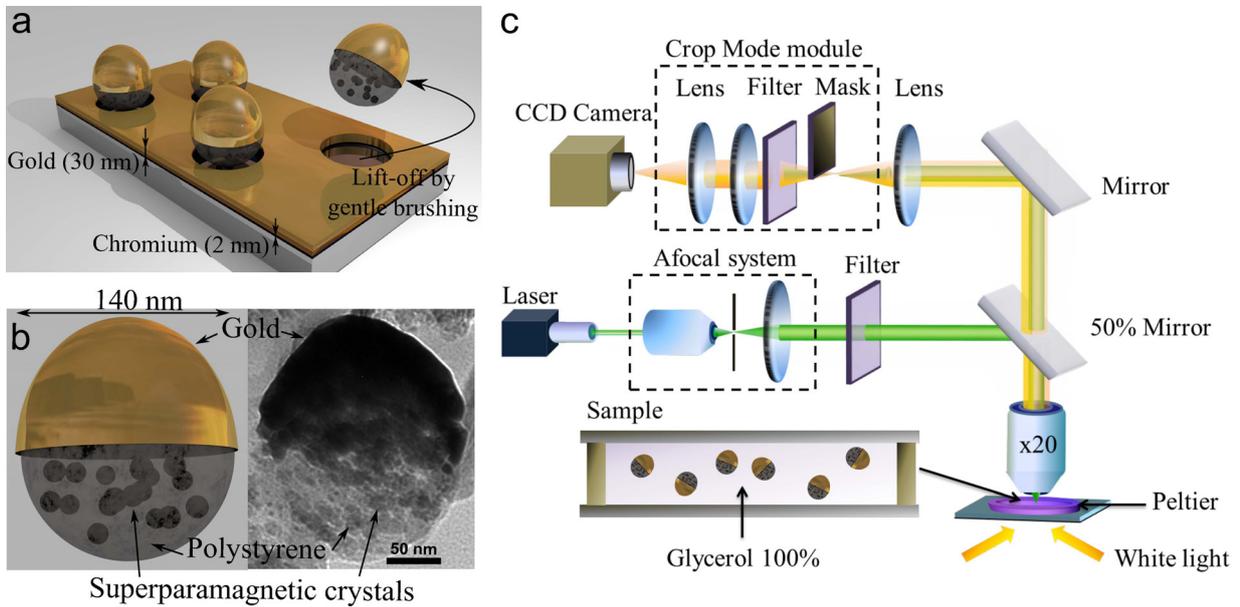

*Figure 1: (a) Schematics of the nanocrescent fabrication by nanosphere lithography. (b) Schematics and TEM image of an individual nanocrescent. (c) Experimental set-up composed of a standard transmission dark-field microscope equipped with a Peltier module. The 150 mW Yag laser emitting at 532 nm is spatially filtered and focused on the sample for photothermic experiments. The CCD camera is equipped with a crop-mode module.*

Figure 1c shows the experimental setup used to measure the RSCS of the nanocrescents and to induce hyperthermia. The sample is observed using a microscope equipped with a standard transmission dark-field configuration with standard white-source illumination (metal halide). The EM-CCD camera (Andor Ixon) is used in a cropped sensor-mode configuration to reach high-frequency acquisition rates of up to 400 Hz. The observation area is defined by a mask positioned in the image plane. The temperature of the sample is controlled by a Peltier module and measured by a thermocouple. Photothermal experiments are performed using a continuous 150 mW Yag laser emitting at 532 nm and focused on the sample.

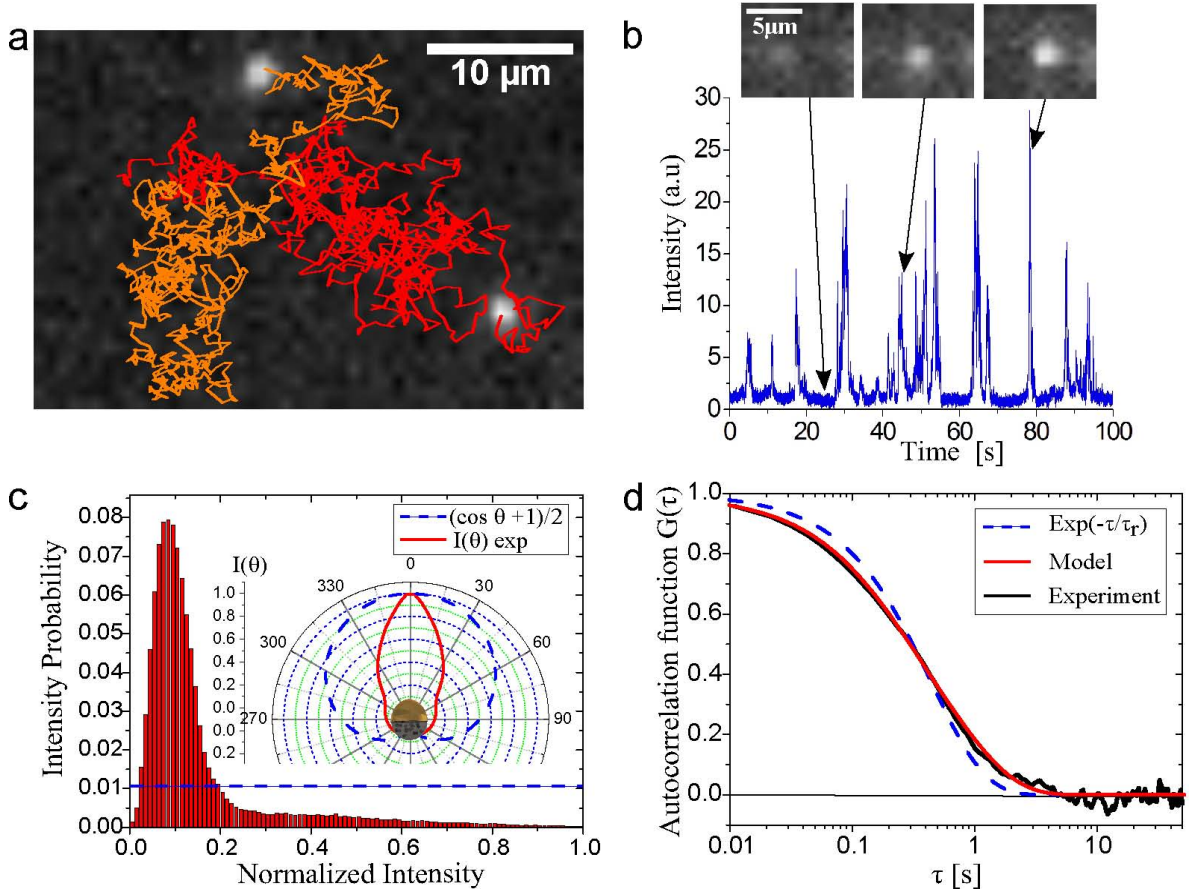

*Figure 2: (a) Brownian trajectories of two nanocrescents. (b) A typical fluctuating intensity I(t) recorded for an individual nanocrescent (insets: associated images on the camera). (c) Histogram of normalized intensities (red bars) and theoretical histogram for a particle having a cosine angle intensity function I(θ) (dashed line). In the inset, the experimental angle intensity function I(θ) (solid line) is compared to a simple cosine function (dashed line). (d) Comparison of the experimental autocorrelation function (black line) with a simple mono-exponential function given by the cosine emission (dashed blue line) and the expression given in eq. (3) with the fitted emission distribution (solid red line).*

To measure the local temperature using RSCS, it is necessary to evaluate the hydrodynamic volume $V_h$ of the nanocrescents (see eq. (4)). This is performed using translational particle diffusion measurements. According to the Stokes-Einstein relation, the mean square displacement is given by $<\Delta r^2(\tau)> = 4D_t\tau$ where $\tau$ is the time elapsed and $D_t$ is the translational diffusion constant. Assuming a spherical shape for the nanocrescents, $D_t(T) = k_BT/3\pi\eta(T)d_h$, where $d_h=(6V_h/\pi)^{1/3}$ is the hydrodynamical diameter. $<\Delta r^2(\tau)>$ is measured at room temperature, using a free particle-tracking algorithm developed by the MOSAIC Group.[37] Figure 2a shows two typical Brownian trajectories of nanocrescents tracked with this algorithm. Averaging over 24 nanocrescents, we found $d_h = 168 \pm 40$ nm, which is in good agreement with the geometric size distribution observed in TEM images.

We now focus on the rotational Brownian dynamics of the nanocrescents. Figure 2b shows a typical fluctuating signal $I(t)$ recorded for an individual nanocrescent. The rotational diffusion causes this signal to blink erratically because of its anisotropic optical response. Sharp maxima alternating with longer low-intensity periods are present. Assuming that each rotational configuration is equiprobable during Brownian motion, we conclude that there are few orientations for which a nanocrescent illuminates the camera. The angular scattering intensity $I(\Omega)\equiv I(\theta)$ thus possesses one or several sharp maxima. Previous studies have shown that gold nanocrescents deposited on transparent dielectric nano-beads possess a dipolar angular scattering intensity with two symmetric maxima along their symmetry axis.[36] We therefore compare the scattering intensities of the nanocrescents when they are oriented with their gold-coated side up or down. The nanocrescents are removed from the glass slide by curing and peeling a casted poly(dimethylsiloxane) (PDMS) polymer film to preserve their orientation.[38] The emission collected from the uncoated side ($\theta = \pi$) is almost completely dampened compared with that of the gold side ($\theta = 0$), suggesting strong absorption by the magnetic nanocrystals. Thus, the expected dipolar symmetry is modified into an unidirectional emission lobe; *i.e.*, $I(\theta)$ presents only one maximum at $\theta = 0$.

More details on the shape of this emission lobe can be obtained from the histogram of the intensity curve $I(t)$ using the equiprobability of orientation of the nanocrescents.[30] Figure 2c shows a typical histogram for a rotating nanocrescent and (inset) the associated emission lobe profile $I(\theta)$ in polar coordinates (solid line). The high probability of recording low intensities in the histogram implies strong emission anisotropy in the angular scattering function. For comparison, the case of a cosine emission profile defined by $I(\theta) \propto \cos\theta + 1$ is also shown (dashed line); it would result in an equiprobable distribution. In the following discussion, we use the experimental function $I(\theta)$ deduced from this procedure to fit the autocorrelation function $G(\tau)$.

Figure 2d shows the autocorrelation function $G(\tau)$ of the measured intensity $I(t)$. It is a decreasing function that becomes zero after a characteristic time of the order of $\tau_r$. This experimental function is fitted both with the theoretical expression given by equation (3) using the experimental profile of $I(\theta)$ (solid line) and with the simple mono-exponential function associated with the cosine approximation (dashed line). The mono-exponential fit does not accurately reproduce the experimental shape of $G(\tau)$; in particular, the decaying slope is too steep. Conversely, the fit using the experimental emission profile accurately reproduces the autocorrelation curve. Moreover, the fitting values obtained for $\tau_r$ using the complete theoretical expression given in equation (3) are approximately 35-40% larger than the ones obtained from the mono-exponential approximation. In the following experiments, all the autocorrelation functions will therefore be analyzed using the full theoretical expression.

We now focus on temperature measurements using RSCS. We control the temperature of the sample using a Peltier module and measure the autocorrelation functions, from which we deduce the temperature of the nanocrescent. We perform a series of temperature measurements on the same single nanocrescent for various temperatures up to 41°C.

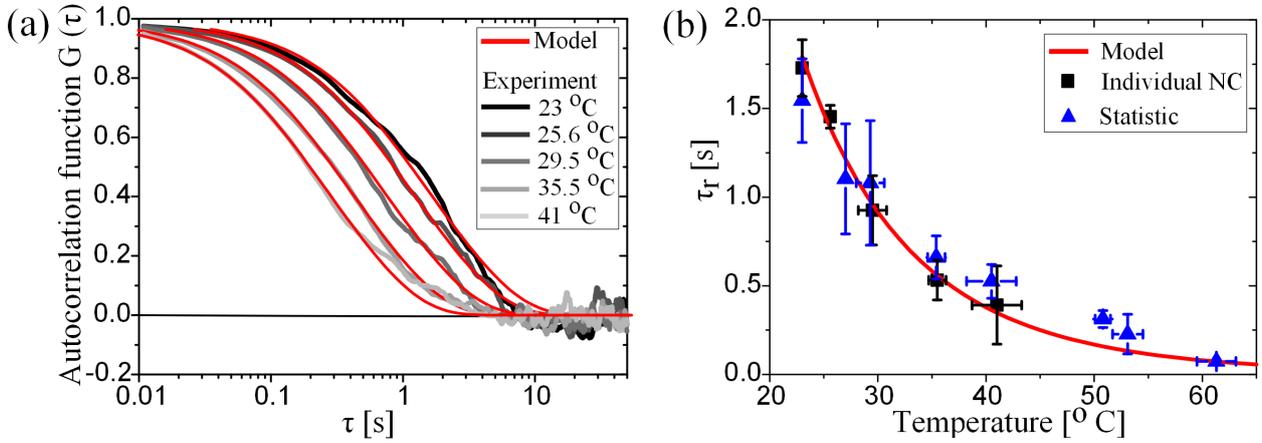

*Figure 3: External control of the temperature using a Peltier module. (a) Autocorrelation functions for an individual nanocrescent in glycerol at different temperatures. Each function is fitted according to eq. (3) (red line) in order to deduce the corresponding $\tau_r$ value. (b) Obtained relaxation times $\tau_r$ as a function of the temperature for both the study on an individual nanocrescent (black scares) and the statistical study (blue triangles). The theoretical $\tau_r(T)$ function is also plotted according to eq. (4) (red line).*

Figure 3a presents the obtained autocorrelation functions and the corresponding fitting curves for this individual particle at various temperatures. Figure 3b shows the deduced relaxation time $\tau_r$ (squares) obtained from the fit of the autocorrelation curves as a function of the reference temperature given by a thermocouple. As expected, the correlation time decreases when the temperature increases. For high temperatures, it is difficult to maintain the same particle in the field of the camera because of its increased translational motion. Hence, we have performed the same type of measurements but have averaged them over several nanocrescents (typically 10 particles) for temperatures up to 61°C (triangles). The deduced relaxation times $\tau_r$ as a function of the temperature match those obtained using a single particle but have higher uncertainties because of the nanoparticle size dispersion. These experimental data are compared in Figure 3b (solid line) with the theoretical curve for $\tau_r(T)$, which was deduced according to eq. (4) by replacing $V_h$ with its mean value, as given by the translational diffusion measurements. The strong temperature dependence of the glycerol viscosity is taken into account using the data given in ref. 39. The theoretical predictions are in good quantitative agreement with the experimental results. This demonstrates the relevance of the RSCS technique performed with nanocrescents to measure accurately the local temperature (with a potential precision of approximately 1°C uncertainty).

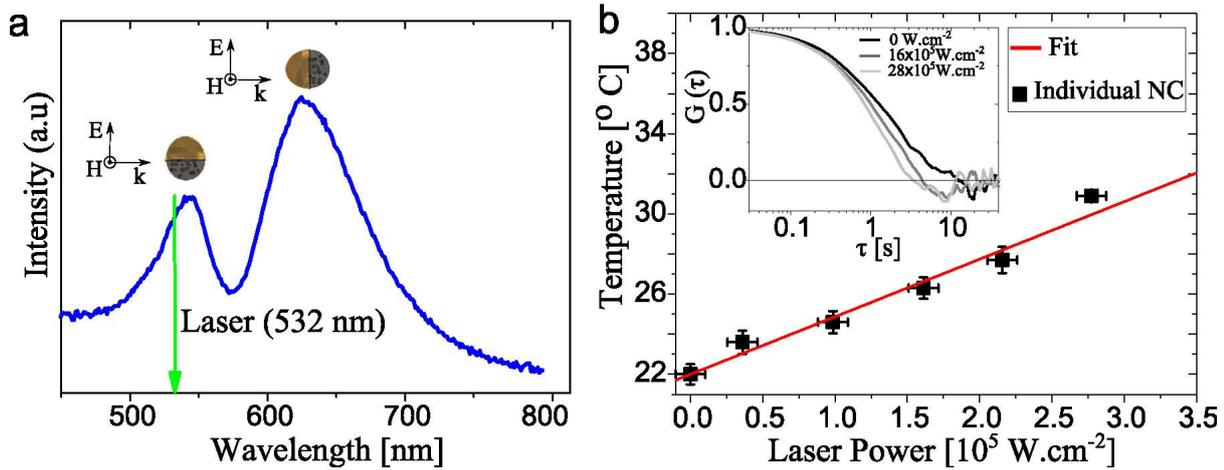

*Figure 4: (a) A typical scattering spectrum of a nanocrescent embedded in PDMS. The laser wavelength is also indicated; it corresponds to the plasmon resonance in the green of the considered nanocrescent. (b) Evaluation of the local temperature of an individual nanocrescent's surroundings for different incident laser intensities. Results are fitted according to eq. (5) in order to deduce the absorption cross-section σ of the considered nanocrescent at the laser wavelength. The inset shows the experimental autocorrelation functions obtained for three intensities of the incident laser.*

We now focus on the local heating properties of a nanocrescent. To induce an efficient photothermic effect with a nanocrescent, it is necessary to illuminate the nanocrescent at the plasmon resonance wavelength. A spectrometer is adapted for the experimental setup described in Figure 1c to measure the scattering spectra of individual nanocrescents embedded in a PDMS slab. A typical spectrum is presented in Figure 4a. Because of their anisotropic shape, nanocrescents possess two strong resonances that can be attributed to a longitudinal electric excitation for the peak centered on 550 nm and to a transverse excitation for the red-shifted peak (as schematically indicated in the figure).[36] The two peak positions are defined with a precision of approximately 20 nm, depending on the nanoparticle size and shape in the colloidal solution.

To heat the nanocrescent, we focused a laser at 532 nm, corresponding to the wavelength of the green plasmon resonance (as indicated in Figure 4a). The strong nanocrescent absorption at the plasmon resonance induces an inhomogeneous temperature distribution because glycerol does not absorb light at this wavelength. Using RSCS, we evaluated the local temperature increase of an individual nanocrescent as a function of the laser power. Figure 4b shows the measured local temperature reached by the nanoparticle as a function of the incident laser power, as deduced from the fit of the autocorrelation functions presented in the inset image. The temperature increases linearly with the laser power as expected. The measured slope is approximately 3.3±0.2 °C/$10^5$ W·$cm^{-2}$. The local temperature increases up to approximately 10°C in the experiment.

The temperature increase is, in first approximation, proportional to the heat dissipation of the nanocrescent. The measured temperature using RSCS is directly related to the nanoparticle temperature. However, the exact relation is not straightforward because of the anisotropy of the nanocrescent and the highly inhomogeneous temperature profile. In equation (5), we use this temperature measurement to estimate the absorption cross-section of the nanocrescent $\sigma_{abs}$. The rotational dynamics are expected to depend on the local temperature of the surrounding medium within a characteristic distance of the hydrodynamic radius. The heat dissipation rate is given by the product of the absorption cross-section $\sigma_{abs}$ and the incident laser power $P$. If we consider the nanocrescents as spherical and homogeneous nanoparticles, the temperature increase $\Delta T(r)$ at $r > R$ outside the particle of radius R is given by:[21]

$$\Delta T(r) = \frac{P\sigma_{abs}}{4\pi k r} \quad \text{with} \quad r > R \qquad (5)$$

where $k$ is the thermal conductivity of the environment.

In the case of glycerol, $k_{gly}$ = 0.28 W.m$^{-1}$.K$^{-1}$. Considering the experimentally measured temperature at the hydrodynamical radius, we obtain an absorption cross-section of $\sigma_{abs.}$ = 9.6x10$^{-12}$ cm$^2$ at the laser wavelength, which corresponds to about 5% of the actual geometric cross-section. It is interesting to compare this result with Mie calculations for nano-shells that give an absorption cross-section similar to the geometric one when excited at resonance. In our case, the smaller result probably originates from the detuning between the laser wavelength and the plasmon resonance maximum. Note that RSCS is particularly well suited to induce the photothermic effect on a single nanoparticle. Because of the very small volume of surrounding medium involved in performing the measurement, the relative temperature increase is high and therefore very sensitive to the temperature increase of the nanoparticle.

In conclusion, we have shown that nanocrescents can be used as nano-thermometers and nano-heaters. Temperature measurements using RSCS can be performed on a single nanoparticle with a precision of approximately one degree. Photothermal effects and local temperature increases can be measured accurately and remotely. This technique should also find applications in nano-rheology. Independent observations of both translational and rotational Brownian diffusions could be of great interest for the study of complex and multiscale systems such as heterogeneous fluids. In such systems, the rotational diffusion, affected by the only particle vicinity, is not necessarily correlated with the translational motion, which probes a larger volume of the environment. In addition, the ability of nanocrescents to act as nano-sources makes them promising tools with which to perform active nano-rheology, with local temperature control, in complex materials and, in particular, in living systems.

**Acknowledgement**

We thank S. Grésillon, O. Loison, E. Bossy, M. Devaud and C. Boccara for fruitful discussions; X.Z. Xu for TEM measurements; F. Monti for his valuable advice and access to the cleanroom; and F. Quinlan-Pluck for helpful comments on this letter. We acknowledge the financial support of the AXA Research Fund, the French National Research Agency and French Program "Investments for the Future" (LABEX WIFI).